\RequirePackage{fix-cm}
\documentclass[twocolumn,epjc3]{svjour3}  
\smartqed  
\RequirePackage{subfig,graphicx,epsfig,amsmath,amssymb}

\journalname{Eur. Phys. J. C}
\begin{document}

\title{Holographic Screening Length in a Hot Plasma of Two Sphere
}


\author{A. Nata Atmaja\thanksref{e1,addr1,addr2}
        \and
        H. Abu Kassim\thanksref{e2,addr1}
        \and
        N. Yusof\thanksref{e3,addr1}
}

\thankstext{e1}{e-mail: ardian\_n\_a@um.edu.my}
\thankstext{e2}{e-mail: hasanak@um.edu.my}
\thankstext{e3}{e-mail: norhaslizay@um.edu.my}


\institute{Quantum Science Centre, Department of Physics\\
Faculty of Science, University of Malaya\\
50603 Kuala Lumpur, Malaysia. \label{addr1}
           \and
           Research Center for Physics \\
Indonesian Institute of Sciences (LIPI)\\
Kompleks PUSPITEK Serpong\\
Tangerang 15310, Indonesia.\\ \label{addr2}
}

\date{Received: date / Accepted: date}

\maketitle

\begin{abstract}
We study the screening length of a quark-antiquark pair moving in a hot plasma living in two sphere $S^2$ manifold using AdS/CFT correspondence where the background metric is four dimensional Schwarzschild-AdS black hole. The geodesic solution of the string ends at the boundary is given by a stationary motion in the equatorial plane as such the separation length $L$ of quark-antiquark pair is parallel to the angular velocity $\omega$. The screening length and the bound energy are computed numerically using Mathematica. We find that the plots are bounded from below by some functions related to the momentum transfer $P_c$ of the drag force configuration. We compare the result by computing the screening length in the quark-antiquark reference frame where the gravity dual are ``Boost-AdS'' and Kerr-AdS black holes. Finding relations of the parameters of both black holes, we argue that the relation between mass parameters $M_{Sch}$ of the Schwarzschild-AdS black hole and $M_{Kerr}$ of the Kerr-AdS black hole in high temperature is given by $M_{Kerr}=M_{Sch}(1-a^2l^2)^{3/2}$, where $a$ is the angular momentum parameter.
\end{abstract}

\section{Introduction}
\label{intro}
One of important signatures of Quark Gluon Plasma produced by heavy ion collision's experiment at RHIC and the new LHC is the suppression of $J/\psi$ mesons production, the $c\bar{c}$ pair. This phenomena is understood qualitatively when temperature of the plasma is larger than the Hagedorn temperature where the potential interaction of $c\bar{c}$ pair would not able to hold them anymore so that $J/\psi$ mesons will dissociate and be screened in the Quark Gluon Plasma\cite{BNL-38344}. The screening potential of $c\bar{c}$ pair depends on a separation length $L$ between $c$ and $\bar{c}$, which could have a value up to some maximum length $L_{max}$, called the screening length, where beyond this length the screening potential becomes flat.

In string theory prescription of AdS/CFT correspondence a heavy quark-antiquark pair, described by a Wilson loop in gauge theory, is defined as a fundamental string where both ends attached on the probe brane at the boundary with different orientation of the electric fields on the ends of string represent a source for the quark-antiquark pair~\cite{Callan:1997kz}. Evaluating the Wilson loop will tell us information about the screening length dependent of the quark-antiquark potential. The procedure to evaluate the Wilson loop goes by extremizing the action of corresponding gravity dual theory as shown in~\cite{hep-th/9803002,hep-th/9803001} for zero temperature case, and~\cite{hep-th/9803135,Brandhuber:1998bs} for finite temperature case.

A calculation on the screening length for a moving quark-antiquark pair at finite temperature four dimensional $\mathcal{N}=4$ Supersymmetric Yang-Mills theory from AdS/CFT correspondence was worked out in~\cite{hep-ph/0607062}. It was then generalized to arbitrary dimension for conformal and non-conformal gauge theories in~\cite{hep-th/0607233}. The calculations there were done by going to reference frame of the moving quark-antiquark pair or explicitly by boosting the background metric to the direction of quark-antiquark pair's velocity. A different approach was done in~\cite{hep-th/0607089} by going to the reference frame of the plasma and furthermore they also calculated the energy from both reference frames. It was found that the screening length is scaled by some power of a boost factor $(1-v^2)$, where $v$ is velocity of the plasma, depending on the dimension of the black hole backgrounds~\cite{hep-th/0607233}. However, the computation of scaling factor is valid in ultra-relativistic limit and disagrees with numerical fitting that was done in~\cite{hep-th/0607089}. For the case considered in this article both scalings may coincide in which the background metric is the four dimensional Schwarzschild-AdS black hole\footnote{The screening length of five dimensional Schwarzschild-AdS Black Hole in Poincare coordinates goes like $L_s\propto(1-v^2)^{1/4}$ as proposed in~\cite{hep-ph/0607062} while in~\cite{hep-th/0607089} it was numerically fitted to $L_s\propto(1-v^2)^{1/3}$. In ultra-relativistic limit, it was derived in~\cite{hep-th/0607233} to be dimensionally dependent in which for the CFT theories $L_s\propto(1-v^2)^{1/d}$, where $d+1$ is dimension of the black hole background.}.

Most of the calculations about the screening length in the literature were done for the case of non-rotating plasma. However in a more realistic scenario the plasma might have some angular momentum such as the Quark Gluon Plasma produced by heavy ion collisions in the RHIC and the LHC. It is natural to think that the peripheral collisions of two nuclei would produce angular momentum to the plasma~\cite{Becattini:2007sr,Abreu:2007kv}. Although the amount of angular momentum left into the resulting plasma is small compared to the initial angular momentum of the two nuclei it is expected that the angular momentum fraction increases as the increasing of collision energy. Phenomenologically the angular momentum might be present in the form of rotation or shearing in the plasma~\cite{Csernai:2011qq,Wang:2013xtp}. Holographically this can be realized by considering asymptotically AdS rotating black holes where topology of the event horizon is spherical or planar respectively. In this article we consider a rotating plasma where the corresponding black hole background is four dimensional Kerr-AdS black hole, with event horizon is two sphere $S^2$, which is more advantageous in improving the chemical potential calculation in the plasma compared to the planar event horizon~\cite{McInnes:2014haa}.

The screening length calculation in the four dimensional Kerr-AdS black hole background is tedious because we have write the metric in asymptotically ``canonical'' AdS (AAdS) coordinates which is very complicated~\cite{Gibbons:2004ai}. As we have learned from the non-rotating case, we could also compute the screening length by going to the reference frame of the plasma as such the quark-antiquark pair is rotating and the background would be four dimensional Schwarzschild-AdS black hole. We therefore start from the screening length calculation in the Schwarzschild-AdS black hole and then compare it with the calculation in the Kerr-AdS black hole, which turn out to be of a special case.

In more detail, we first compute the screening length of a heavy quark-antiquark, $q\bar{q}$, pair moving in the plasma that lives in a compact space of two dimensional sphere $S^2$ where the corresponding background metric is four dimensional Schwarzschild-AdS black hole in the Global coordinates. We further would like to make comparison with the computation in the reference frame of the $q\bar{q}$ pair where the plasma is rotating hence the background metric is a stationary rotating black hole. Unlike the Poincare coordinates case, generally the rotating AdS black holes in the Global coordinates can not be obtained simply by boosting procedure as in~\cite{hep-th/0607233}. However the solution for rotating black holes in the Global coordinates by a different procedure are available and its known as Kerr-AdS black holes. For particular case in this article we will also compare the screening length between the black hole called ``Boost-AdS'' black hole, where the boosting procedure is doable, and the well-known solution Kerr-AdS black hole.

In the Poincare coordinates case we can compute the screening length for arbitrary angle between the separation length of quark-antiquark pair and its velocity direction or the hot wind plasma direction. Unfortunately the situation is quite restricted for the Global coordinates case. Main reason is because the geodesic of heavy quarks at the boundary must follow the great circle solutions. Since we want to keep the screening length to be fixed, both quark and antiquark must stay on the same great circle plane for stationary motion. In that way, the only possible angle is when the separation length of quark-antiquark pair is parallel with its angular velocity. Furthermore, using $SO(3)$ symmetry we can rotate arbitrary great circle planes to the equatorial plane. This turn out to be a benefit in the screening length calculation of the Kerr-AdS black hole where the explicit expression of the background metric in AAdS coordinates is simple.

In the next section \ref{section 2}, we will compute the screening length in the reference frame of the plasma. We plot numerically the screening length for various fixed angular velocities as a function of momentum transfer of angular coordinate $\phi$ along the string, $\pi^\sigma_\phi\propto P$. We also compute the bounded energy density as a function separation length. In section \ref{section 3}, we proceed the computation in the reference frame of the quark-antiquark pair. Unlike the Poincare case we have more than one background metrics which are ``Boost-AdS'' black hole and Kerr-AdS black hole. We then try to compare the results by finding relations between parameters of these black holes. Therefore we can plot numerically the screening length as a function of angular velocity parameter for these black holes using the aforementioned relations. The last section \ref{section 4} is discussion and conclusion of the results of the previous sections.

\section{Screening Length in Plasma Reference Frame}
\label{section 2}
In the plasma reference frame the background metric is static and is given by four dimensional Schwarzschild-AdS black hole in the Global coordinates written as
\begin{eqnarray}
 ds^2&=&-r^2 h(r) dt^2 + {1\over r^2 h(r)}dr^2+r^2\left(d\theta^2 +\sin^2\theta~ d\phi^2\right),\label{AdS-S metric}\nonumber \\
h(r)&=&l^2+{1\over r^2}-{2 M_{Sch} \over r^3},\ T_{Sch}={1\over 4\pi}\left({1\over r_H}+3 r_H l^2\right),
\end{eqnarray}
with $r_H\leq r< \infty$, $0\leq \theta\leq \pi$, and $0\leq \phi <2\pi$. Here $l$ is the curvature radius of AdS space, $M_{Sch}$ is proportional to the mass of the Schwarzschild-AdS black hole, $T_{Sch}$ is the Hawking temperature, and $r_H$ is the event horizon given by the most positive real roots of $h(r)$, $h(r_H)=0$. This Schwarzschild-AdS black hole has a minimum temperature and two branches of high temperature. We choose the branch for $r_H l>1/\sqrt{3}$ or $M_{Sch}l>{2\over 3\sqrt{3}}$ which is favored thermodynamically~\cite{Hemming:2007yq}. In this metric the corresponding plasma lives in the boundary of $AdS_4$ which is the three dimensional Einstein static universe, where the spatial manifold is two sphere $S^2$.

Classical solution of a string is obtained by solving the equation of motion from the Nambu-Goto action under the background (\ref{AdS-S metric}),
\begin{align}
 S=- T_0\int d\sigma^2 \sqrt{-\det g_{\alpha\beta}}, \nonumber\\
 \ \ \ \ \ \ \ g_{\alpha\beta}\equiv G_{\mu\nu}\partial_{\alpha}X^\mu \partial_{\beta}X^\nu,\ \ \ \ \ \ \ T_0={1\over 2\pi\alpha'},\label{Nambu-Goto}
\end{align}
where $\sigma^{\alpha}\equiv(\tau,\sigma)$ is the worldsheet coordinates, $X^\mu(\sigma^\alpha)$ is the space-time coordinates where the string worldsheet is embedded, and $G_{\mu\nu}$ is the space-time metric (\ref{AdS-S metric}). The equation of motion derived from action (\ref{Nambu-Goto}) is simply written as
\begin{align}
 \ \ \ \nabla_\alpha P^\alpha_\mu&=0,\ \ \ P^\alpha_\mu\equiv{\pi^\alpha_\mu \over \sqrt{-g}},\ \ \ \pi^\alpha_\mu\equiv{\delta S \over \delta\partial_\alpha X^\mu},
\end{align}
where $\pi^\alpha_\mu$ is the cannonical worldsheet momentum with $g=\det g_{\alpha\beta}$.

In deriving the equation of motion we will use the gauge $\tau=t$ while for $\sigma$ will be determined later for convenient. As it was explained in the Introduction we will consider the solution on equatorial plane $\theta=\pi/2$. Taking an ansatz
\begin{align}
 \phi(\sigma^\alpha)&\equiv \omega \tau+\phi(\sigma),\label{ansatz phi}\\ 
 r(\sigma^\alpha)&\equiv r(\sigma),\label{ansatz r}
\end{align}
 for a moving quark-antiquark with angular velocity $\omega$, the equations of motion are given by:\\
\begin{align}
 \phi'^2\left(r^4h\right)'+&r'^2\left(1-{\omega^2\over h}\right)'\notag\\
 &-{2r'\sqrt{-g}}{\partial \over \partial\sigma}\left[{r'(h-\omega^2) \over h\sqrt{-g}}\right]=0,\label{EOM r}\\
 \partial_\sigma\pi^\sigma_\phi&\equiv{\partial\over\partial\sigma}\left[{r^4h\over \sqrt{-g}} \phi'\right]=0,\label{EOM phi}
\end{align}
with $\pi^\sigma_\phi$ is constant and $'\equiv {\partial\over\partial\sigma}$. There are two ways to define a gauge for $\sigma$ coordinate. The first one is $\sigma=\phi$ gauge which is actually nice in the quark-antiquark configuration since $r(\phi)$ is a single valued function. However the equation obtained, which is given by equation (\ref{EOM r}), is not simple. The other choice of gauge is $\sigma=r$. One will obtain the equation (\ref{EOM phi}) and it has a reflection symmetry in $\phi$. It yields that for every solution of $\phi(r)$ than its $-\phi(r)$ is also a solution. In the quark-antiquark configuration, taking the $\cup$-shape in $r-\phi$ plane, the $\phi(r)$ is a double valued function. It will turn out that the double valued function $\phi(r)$ in the quark-antiquark pair configuration is obtained by pairing these solutions of $\phi$ and $-\phi$. The equation (\ref{EOM phi}) is rather simple than the former one. Another advantage there is a conserved momentum transfer $\pi^\sigma_\phi$ which can be useful in the discussion of physical properties of the string configuration. Therefore we are going to use the gauge $\sigma=r$ from now on\footnote{The gauge can be taken before deriving the equations of motion or after deriving them. In the later case there are still two equations which can be proved to be equivalent.}.

For quark-antiquark pair, following~\cite{hep-th/0607089}, we take the first solution for $\phi$ that must satisfy conditions
\begin{align}
 \phi(r\to\infty)=L/2,~~~~~ \phi'(r_p)=\infty,\label{conditions}
\end{align}
where $L$ is a dimensionless separation length of the quark-antiquark pair and $r_p$ is the turning point where $r$ takes the minimum value with $r_p> r_H$. Using reflection symmetry, $\phi\to -\phi$, we can set the turning point in the middle such that $\phi(r_p)=0$. In the holographic formulation the quark/antiquark is located at the boundary $r\to\infty$. This could be the source of divergence in the calculation of energy. However, we will see later this divergence can be removed by extracting the the energy of each quark and antiquark in a single string configuration. We will find later it is surprising that the separation length $L$ is finite although the string length, given by formula $L_{string}=2\int^\infty_{r_p} dr\sqrt{{1\over r^2 h}+r^2\phi'^2}$, is infinite. Since the length of string is not in our interest, so we do not have to put an UV cut-off to regularize the results. We carefully choose the values of parameters so that the computed $L$ is smaller than the range of coordinate $\phi$ since the coordinate $\phi$ is periodic with length $2\pi$, $-\pi\leq\phi<\pi$. Throughout this paper all numerical calculations will be expressed in terms of dimension less quantities where the conversion is carried out by the curvature radius $l$, e.g. the angular velocity $\omega$ with dimension $[\mbox{length}]^{-1}$ in dimension less quantity is written as $\omega/l$.
\begin{figure}[th]
\centerline{\includegraphics[width=10cm]{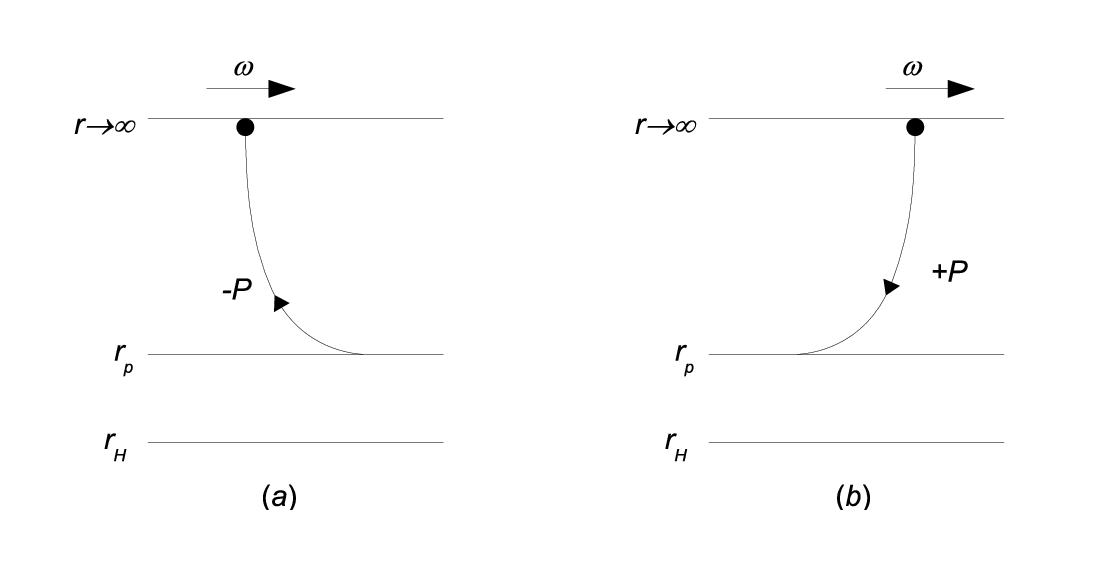}}
\caption{Two different configurations determined by the sign in $P$. The positive sign solution in (b) while the negative sign solution in (a). \label{draging}}
\end{figure}

Solutions to the equation (\ref{EOM phi}) are given by
\begin{equation}
 \phi'=\pm {P\over r^2 h}\sqrt{h -\omega^2\over r^4h-P^2},\label{sol r AdS-S}
\end{equation}
where now $'\equiv{\partial\over\partial r}$ and $P\propto\pi^\sigma_\phi$ is a constant denoting amount of $\phi$-component of momentum transfer on the string. In our convention $P>0$, while $P=0$ would correspond to a straight string configuration, and the positive sign in (\ref{sol r AdS-S}) is proportional to the energy flows from the boundary down to $r=r_p$ while the negative sign is the opposite. The quark-antiquark pair can be built out of these two solutions correspond to configurations (a) and (b) as shown in Figure \ref{draging}. The negative sign solution may seems to be unphysical, according to the drag force analysis \cite{Gubser:2006bz,Herzog:2006gh}, however this is not the case since the energy transfer is not coming from the horizon but from the turning point $r_p>r_h$. It is necessary to take these two configurations in order for the string of quark-antiquark pair to join up at $r=r_p$ as such the energy transfer, ${dE\over dt}\equiv -\pi^\sigma_t=-T_0P\omega$, is coming from one end of the string at the boundary down to the turning point back again to the other end of the string on the boundary. The energy transfer is in the opposite direction to the angular velocity as depicted in Figure \ref{draging}. Unlike the perpendicular case in \cite{hep-th/0607089} the constant force, ${dp_\phi \over dt}=\pi^\sigma_\phi=-T_0P$, is non-vanishing by the non-zero force at the turning point since the string is parallel to the $\phi$-axis. In more detail the double valued $\phi$ in our setup is defined as follows
\begin{eqnarray}
\label{r variation}
 {d\phi\over dr}=\left\{\begin{array}{ll}
           {P \over r^2 h}\sqrt{h -\omega^2 \over r^4h-P^2}~~~~~ & ,0\leq\phi\leq L/2 \\
           -{P \over r^2 h}\sqrt{h -\omega^2 \over r^4h-P^2}~~~~~ & ,-L/2\leq\phi\leq 0  
          \end{array}
\right. .
\end{eqnarray}

\begin{figure}[ht]
\centerline{\includegraphics[width=9cm]{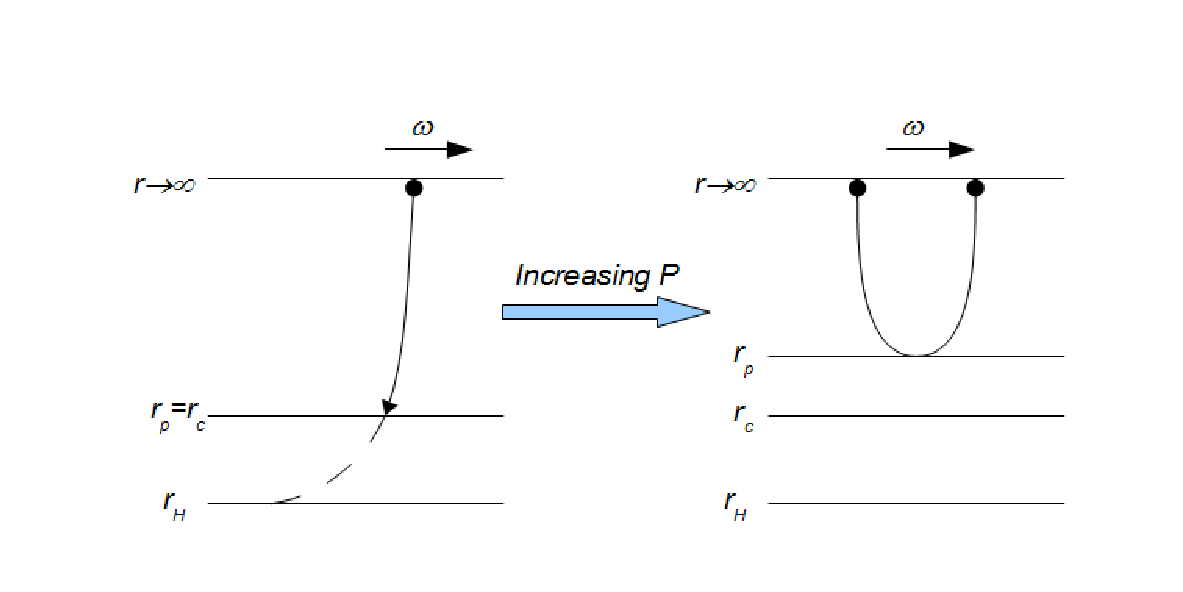}}
\caption{Picture on the left shows at critical momentum transfer $P^2=P_c^2=\omega^2 r_c^4$ the string describe a single quark while picture on the right is if we increase $P$ such that $P^2>\omega^2 r_c^4$. \label{string conf}}
\end{figure}

We solve the solution of (\ref{sol r AdS-S}) for positive sign which corresponds to configuration (b) in Figure \ref{draging} while the other solution can be obtained by changing $P\to-P$. At $r=r_p$, we must have $r_p^4 h(r_p)-P^2=0$ which implies that $h(r_p)>0$ or $r_p>r_H$ for $P>0$. There is a critical radius $r_c$ defined as $h(r_c)=\omega^2$ and it is also that $r_c>r_H$ for $\omega\neq 0$. Solving (\ref{sol r AdS-S}) in terms of integral formula for $r_p>r_c$ then the integral in coordinate $r$ is cut-off below at $r=r_p$ and string configuration for the quark-antiquark pair is formed. On the other hand if $r_p<r_c$ the integral is also cut-off below at $r=r_c$ however this solution requires $P=0$ at $r=r_c$ since here the string is perpendicular to the $\phi-$axis, which is a similar situation as in \cite{hep-th/0607089}, therefore the physical configuration is a straight string. We may ignore the configuration for $r_p=r_c$ where the condition (\ref{conditions}) could not be satisfied and so the string configuration physically would be similar to the drag force configuration, curved moving string. Thus the string configuration for quark-antiquark pair requires $r_p>r_c$ or $P^2>P_c^2=\omega^2 r_c^4$. Notice that this $r_c$ equals to $r_{Sch}$ as appears in the drag force computation hence we interpret the constant $\pi^\sigma_\phi$ in~\cite{NataAtmaja:2010hd,Chunlen:2010ar} equals to $P_c$ because both are given by the same formula (\ref{EOM phi}) with $r_p=r_c$. So for some fixed angular velocity $\omega$, or $r_c$, the string configuration is characterized by the amount of momentum transfer $P$ on the string with $P^2\geq P_c^2$. At $P^2=P_c^2$, the string tends to represent a single quark and if we increase the momentum transfer $P$ the quark-antiquark pair is formed. Figure \ref{string conf} shows schematic pictures of string configuration describing quark-antiquark pair and drag force of a single quark.

\begin{figure}[t]
\centerline{\includegraphics[width=8cm]{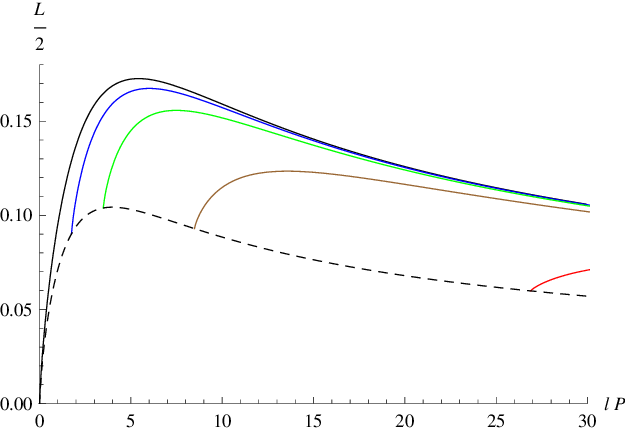}}
\caption{Plot of dimensionless separation length for various $\omega/ l=0,0.25,0.45,0.75,0.95$, with fixed $M_{Sch}l=10$, as functions of $Pl$ where the corresponding colors are black, blue, green, brown, and red. The dashed black line is the plot of bounded momentum transfer obtained by substituting $\omega^2= P^2/r_p^4$, for which $r_c=r_p$, into formula (\protect\ref{integral form}). In this plot the quark-antiquark pair can not be formed instead the string becomes a single quark configuration or the drag force configuration.} \label{Plots AdS-S}
\end{figure}

From equation (\ref{sol r AdS-S}), we obtain an integral formula for computing the separation length as below\footnote{Our calculation here is in the same footing as in~\cite{hep-ph/0607062,Natsuume:2007vc,Argyres:2006vs} for parallel case, $\theta=0$.}
\begin{align}
\label{integral form}
 \int_0^{L/2} d\phi={L\over 2}=\int_{r_p}^\infty dr {P\over r^2 h}\sqrt{h-\omega^2 \over r^4 h-P^2}.
\end{align}
The integral above can not be solved analytically so we are going to plot the integral numerically. The numerical solution of the separation length for various value of $\omega/l$ is shown in Figure \ref{Plots AdS-S}.
As one can see, we produce similar profile of screening length $L_s$ as in the Poincare case of~\cite{hep-th/0607089} except that the momentum transfer is bounded from below at $P^2=P_c^2\equiv\omega^2 r_c^4$, for fixed $\omega$. Therefore our plots in general do not start from $P=0$ which is different from the plots produced in~\cite{hep-th/0607089} instead we start from $P=P_c$. Similar to the perpendicular case, the separation length in Figure \ref{Plots AdS-S} consist of two regions separated by the maximum $L_{max}$, called the screening length, at the point $P=P_{max}$. For $P_c<P<P_{max}$, the string configuration is metastable while for $P\geq P_{max}$ is mostly stable depending on the potential energy of the quark-antiquark pair which we will compute in the next section, for detail discussion can be read in~\cite{hep-ph/0607062,hep-th/0607089}. As one can see we exclude the $P=P_c$ in the string configuration of quark-antiquark pair since the string geometry is not cut-off at $r_p$ instead it can be continued down to $r_H$ thus the integral (\ref{integral form}) should be taken from $r=r_H$ to $r\to\infty$. However the separation length in this case will be infinite, or at least larger than the screening length, because the integrand is divergent near $r\to r_H$ thus this support our analysis that the string configuration of quark-antiquark pair is dissociated, where the quark or antiquark becomes a single string configuration, due to this infinitely large separation length.

\subsection{Drag force}
\label{drag force}
As we mentioned previously the amount of momentum transfer is given by a constant $P$ and the total force is
\begin{align}
 {dp_\phi \over dt}=-T_0 P, \ P>0.
\end{align}
This is denoting the amount of drag force experienced by the quark-antiquark pair moving in the plasma corresponds to Schwarzschild-AdS black hole background (\ref{AdS-S metric}). The amount of momentum transfer $P$, for fixed separation length $L$, is actually depends on angular velocity $\omega$ as we can infer from Figure \ref{Plots AdS-S}. The negative sign shows that this drag force tends to decrease the angular momentum, or the angular velocity, of the quark-antiquark pair. In order to have the pair moves with a constant angular velocity and fixed separation length we must supply an external force at the boundary to overcome the drag force. Another way to see this is by noticing that the equations of motion (\ref{EOM r}) and (\ref{EOM phi}) are subject to the boundary terms
\begin{align}
 \delta S_{bd}=\int d\tau \left.\left[\pi^\sigma_\phi \delta\phi+\pi^\sigma_r \delta r\right]\right|^{L_{string}/2}_{\sigma=-L_{string}/2}.
\end{align}
For an open string where both ends fixed at the boundary $r\to\infty$, we have $\delta r(\pm L_{string}/2)=0$. In case of a fixed the separation length $\delta\phi(L_{string}/2)=\delta\phi(-L_{string}/2)$ the boundary terms then is non-vanishing, $\delta S_{bd}\propto -T_0P$. So in this case we need to impose an external force at the boundary to overcome the non-vanishing boundary terms.

\subsection{Total energy of the string}
We are now interested to compute the total energy of the string. The bounded energy density of the quark-antiquark pair is given by~\cite{Herzog:2006gh}
\begin{align}
 \mathcal{H}_{bound}=2 T_0 {r^4h^2-P^2\omega^2 \over r^2h\sqrt{(h-\omega^2)(r^4h-P^2)}},
\end{align}
As usual the bounded energy is linearly divergent near the boundary, $r\to\infty$, therefore it must be subtracted with the unbounded energy density of independent quark and antiquark moving with constant angular velocity $\omega$ which is, see~\cite{NataAtmaja:2010hd}, 
\begin{align}
 \mathcal{H}_{unbound}=2 T_0 {r^4h^2-P_c^2\omega^2 \over r^2h\sqrt{(h-\omega^2)(r^4h-P_c^2)}}.
\end{align}
This unbounded energy will cancel the boundary divergence of the bounded energy with the cost of producing a new divergence near the horizon. As explained in~\cite{hep-th/0607089}, and also in~\cite{Herzog:2006gh}, this source of divergence coming from the unbounded energy density because of infinite amount of energy flowing down from boundary, supplied by external force, to the horizon to keep the quark and antiquark move with constant angular velocity. Furthermore this also produce some ambiguity in removing this horizon divergence.
\begin{figure*}[ht]
   \subfloat[$\omega/l=0$\label{subfig-1:dummy}]{%
      \includegraphics[width=7cm]{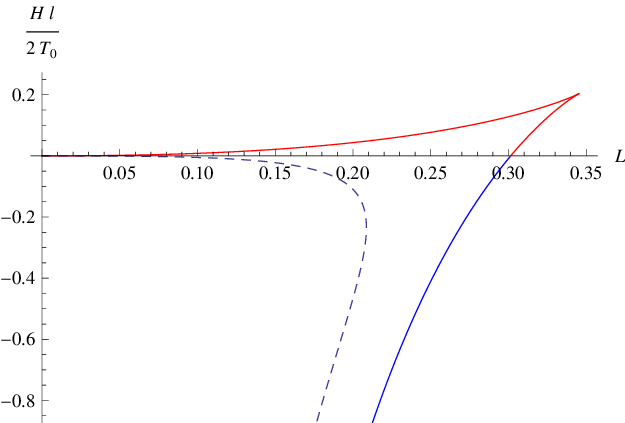}
    }
    \hfill
    \subfloat[$\omega/l=0.25$\label{subfig-2:dummy}]{%
      \includegraphics[width=7cm]{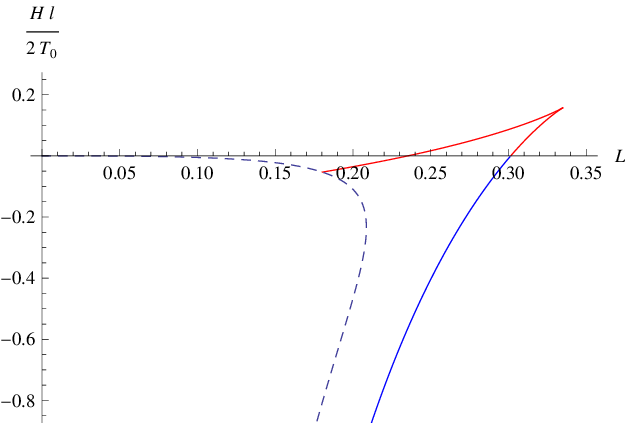}
    }
   \vfill
    \subfloat[$\omega/l=0.45$\label{subfig-3:dummy}]{%
      \includegraphics[width=7cm]{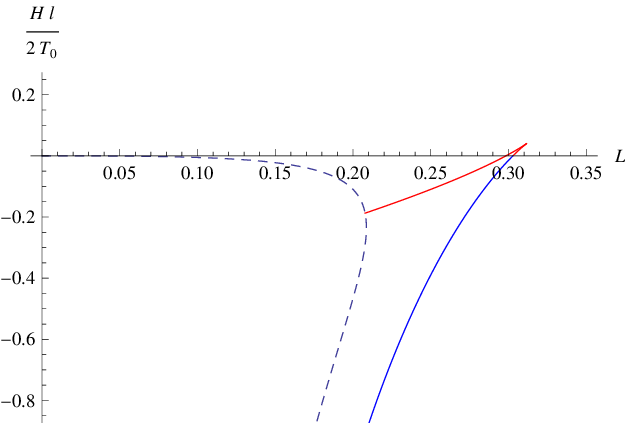}
    }
    \hfill
    \subfloat[$\omega/l=0.75$\label{subfig-4:dummy}]{%
      \includegraphics[width=7cm]{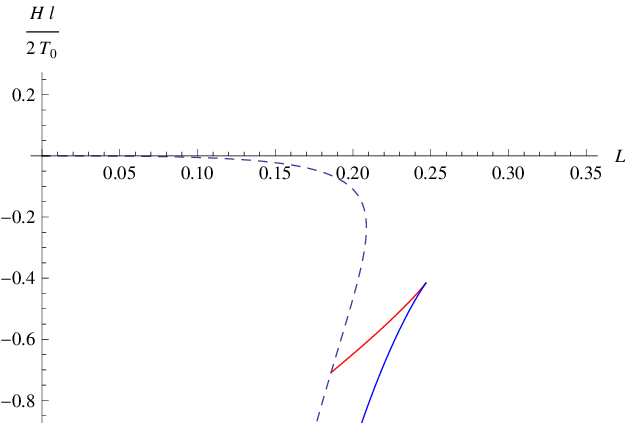}
    }
    \caption{Plots of normalized energy for quark-antiquark pair $Hl/2T_0$ as function of separation length $L$ for various angular velocity $\omega/l$ with fixed $M_{Sch}l=10$. The metastable configuration is colored with red while the stable configuration is colored with blue. The dashed black plot is again the saturated energy $H_{sat}$ where all possible energy on this plot is forbidden for quark-antiquark pair.}
    \label{fig:dummy}
\end{figure*}

Following~\cite{hep-th/0607089}, as a resolution we need to compute the bounded and unbounded energy density in the rest frame of the moving string and this can be done, using the symmetry of the Einstein static universe at the equatorial, by the following boost transformation:
\begin{align}
\label{boost trans}
 t\to\gamma\left(t-{\omega\over l^2} \phi\right),\ \phi\to\gamma\left(\phi-\omega t\right),\notag \\
 \gamma=\left(1-{\omega^2 l^{-2}}\right)^{-1/2},
\end{align}
where $\gamma$ is the boost factor. The resulting three dimensional ``Boost-AdS'' black hole is
\begin{align}
\label{Boost}
 ds^2=&-r^2 h(r) \gamma^2\left(dt-{\omega\over l^2} d\phi\right)^2 + {1\over r^2 h(r)}dr^2\notag\\
 &+r^2 \gamma^2\left(d\phi-\omega dt\right)^2,
\end{align}
where the Hawking temperature is now $T_{Boost}=T_{Sch}/\gamma$. One can check applying this transformation to the four dimensional Schwarzschild-AdS black hole still satisfies the Einstein equation with negative cosmological constant, $R_{\mu\nu}=-3 l^2 g_{\mu\nu}$. The actual boost factor, for arbitrary $\theta$, should depend on $\theta$ and it is given by $\gamma=\left(1-{\omega^2 l^{-2}}\sin^2\theta\right)^{-1/2}$. Unfortunately, using the same boost transformation with this actual boost factor the rotating version of four dimensional Schwarzschild-AdS black hole does not satisfy the Einstein equation $R_{\mu\nu}=-3 l^2 g_{\mu\nu}$, even at the equatorial.

Taking the same gauge as before, $\tau=t$ and $\sigma=r$, the solution for $\phi(r)$ is given by
\begin{align}
  \phi'= {P \gamma \over r^2 h}\sqrt{h -\omega^2 \over r^4 h-P^2},\label{sol r Boost}
\end{align}
where $P$ is again the same $\phi$-component of momentum transfer as in previous static black hole case. One can immediately see from (\ref{sol r Boost}) that separation length will be different from the static case by a boost factor $\gamma$ thus the screening length as well, $L_{Boost}=\gamma L_{Sch}$. In this ``Boost-AdS'' black hole the bounded energy density turns out to be
\begin{align}
 \mathcal{H}_{bound}= 2\gamma T_0 r^2\sqrt{h-\omega^2 \over r^4 h-P^2},~~~~~P^2>\omega^2 r_c^4.
\end{align}
with the integral range $r_p\leq r<\infty$. The corresponding unbounded energy density is simply written as
\begin{align}
 \mathcal{H}_{unbound}= 2\gamma T_0 r^2\sqrt{h-\omega^2 \over r^4 h-P_c^2},~~~~~P_c^2=\omega^2 r_c^4.
\end{align}
As we can see the boundary divergence of the bounded energy is removed by the unbounded energy without introducing the horizon divergence. So, we can define a normalized energy for the quark-antiquark pair
\begin{align}
 H=\int^\infty_{r_p} dr~ \mathcal{H}_{bound}-\int^\infty_{r_H} dr~ \mathcal{H}_{unbound}.
\end{align}
Rewriting the normalized energy $H$ as a function of separation length $L$ instead of $P$ using formula (\ref{integral form}), we plot numerically the normalized energy for various value of angular velocity $\omega$, see Figure \ref{fig:dummy}. For plots (a), (b), and (c) the separation length does not reach its screening length, at the tip of the curve. The characteristic behavior of the normalized energy is similar as in~\cite{hep-th/0607089} except that the normalized energy for each value of angular momentum $\omega$ has a minimum length for metastable configuration, in red plot of Figure \ref{fig:dummy}, given by the dashed black plot. This dashed black plot is produced by the saturated energy of the following function
\begin{align}
  H_{sat}=-\int^{r_c}_{r_H} dr~ \mathcal{H}_{unbound}.
\end{align}

\section{Screening Length in Quark-Antiquark Reference Frame}
\label{section 3}
A different way of computing the screening length holographically is by going to the reference frame of the moving quark-antiquark pair. In this case the metric will be seen as it is rotating at angular velocity proportional to angular velocity of the quark-antiquark pair. In Poincare case, this metric is provided is by boosting the static metric with a procedure described in~\cite{Bhattacharyya:2008jc}. Unfortunately this procedure can not be done in the Global case in particular under the boost transformation (\ref{boost trans}) with the actual boost factor, for arbitrary $\theta$, as explained previously. However the corresponding rotating metric is available in~\cite{Carter:1968ks,Hawking:1998kw} and a procedure to construct the rotating metric from the static metric in the Global case was first given in~\cite{Kim:1998iw} for three dimensional Schwarzschild-AdS black hole to obtain the well-known BTZ black hole using the Newman-Janis procedure~\cite{Newman1965}, which is very unfortunate that we have not found its application to the four dimensional Schwarzschild-AdS black hole in the literature so far. A different procedure for constructing the rotating metric in Global case for arbitrary dimensions was given in~\cite{Gibbons:2004uw}. Here we will not discuss the procedure of how to construct the rotating metric but we just use the resulting metric in various coordinate representations.

\subsection{Kerr-AdS Black Hole in Boyer-Linquist coordinates}
There are many coordinate representations for Kerr-AdS black holes~\cite{Gibbons:2004uw}. The simplest one is using Boyer-Linquist coordinate system. We assume the equatorial plane in the Kerr-AdS black hole is similar to the equatorial plane in the Schwarzschild-AdS black hole by means we identify the both polar angle coordinate-$\theta$ in these two coordinate systems to be equal at the equatorial, $\theta=\pi/2$. The four dimensional Kerr-AdS Black Hole metric in Boyer-Linquist coordinates at equatorial is given by~\cite{Carter:1968ks,Hawking:1998kw}
\begin{eqnarray}
\label{Boyer-Linquist}
 ds^2&=&-{\Delta_r \over r^2}\left(dt-{a\over \Xi}d\phi\right)^2+{r^2 \over \Delta_r}dr^2\notag\\
 &&+{1\over r^2}\left(a dt-{r^2+a^2 \over \Xi}d\phi\right)^2,\notag\\
\Delta_r&=&(r^2+a^2)(1+l^2r^2)-2M_{Kerr}r,\ \Xi=1-a^2l^2
\end{eqnarray}
with the Hawking temperature, equal to the full four dimensional black hole,
\begin{equation}
T_{Kerr}={r_K(3l^2r_K^2+1+a^2l^2-a^2r_K^{-2}) \over 4\pi (r_K^2+a^2)},
\end{equation}
where event horizon $r_K$ is the largest positive root of $\Delta_r$.
Using the gauge, $\tau=t$ and $\sigma=r$, the solution for $\phi(r)$ is now given by
\begin{align}
   \phi'= {P\Xi^2 \over \Delta_r }\sqrt{\Delta_r -a^2 \over \Delta_r-P^2\Xi^2}.\label{sol r Boyer-Linquist}
\end{align}
The momentum transfer $P$ here, although a constant, has a different formula with the one in Schwarzschild-AdS black hole case and this could give us a different separation length function of $P$ compare to the previous one. The formula for momentum transfer $P$ can be extracted from equation (\ref{sol r Boyer-Linquist}) above. One thing we are worried about in this Boyer-Linquist coordinate representation, although simple, the asymptotic behavior of the metric is not $AdS_4$ at the boundary which is different from the Schwarzschild-AdS black hole. This might turn out to give different CFT theory or plasma at the boundary in the context of AdS/CFT correspondence. Therefore we need to use more adequate coordinate system as discussed in the next section.

\subsection{Kerr-AdS Black Hole in Asymptotically-AdS coordinates}
Another representation of Kerr-AdS metric is using AAdS coordinate system. This Kerr-AdS metric in AAdS coordinate system is $AdS_4$ asymptotically, at the boundary $Y\to\infty$, hence is preferred by AdS/CFT correspondence prescription~\cite{Witten98-1,Gubser:1998bc}. In general, to get the full geometry in AAdS coordinate system is quite involved, yet it is still possible to write down the metric for particular case such as at the equatorial plane which again we take to be equal to the equatorial plane of the Schwarzschild-AdS black hole. The four dimensional Kerr-AdS metric in AAdS coordinates at equatorial can be obtained by the following coordinates transformation from the metric (\ref{Boyer-Linquist})~\cite{Hawking:1998kw,Gibbons:2004ai}:
\begin{align}
\label{Kerr-AdS trans}
 \Xi Y^2=r^2+a^2,~~~~~ \Phi=\phi-al^2 t.
\end{align}
The resulting metric is
\begin{eqnarray}
\label{AAdS}
 ds^2&=&-{\Delta_Y \over \Xi^2(\Xi Y^2-a^2)}\left(dt-a d\Phi\right)^2+{\Xi^2 Y^2 \over \Delta_Y}dY^2\notag\\
 &&+{1\over \Xi Y^2-a^2}\left(a(1+l^2Y^2) dt-Y^2 d\Phi\right)^2,\notag\\
\Delta_Y&=&Y^2\Xi^2(1+l^2Y^2)-2M_{Kerr}\sqrt{\Xi Y^2-a^2},\notag\\
\Xi&=&1-a^2l^2,
\end{eqnarray}
where now the Hawking temperature is written as
\begin{equation}
 T_{Kerr}= {Y_K^2+ 3l^2 Y_K^4-a^2(2+5l^2 Y_K^2+3l^4 Y_K^4)\over 4\pi Y_K^2\sqrt{Y_K^2-a^2(1+l^2 Y_K^2)}},
\end{equation}
with even horizon $Y_K$ is the largest positive root of $\Delta_Y$. One can check that it is equal to the Hawking temperature of the Boyer-Linquist case taking that $\Xi Y^2_K=r^2_K+a^2$. Using a gauge, $\tau=t$ and $\sigma=Y$, the solution for $\Phi(Y)$ is given by
\begin{align}
   \Phi'= {P\Xi^2 \over \Delta_Y}{\sqrt{Y \over \Xi Y^2-a^2}}\sqrt{\Delta_Y -a^2\Xi^2(1+l^2Y^2)^2 \over \Delta_Y-P^2\Xi^2}.\label{sol r AAdS}
\end{align}
Again, the momentum transfer $P$ here is different from the Schwarzschild-AdS black hole, and also in the Boyer-Linquist coordinate system, and the exact expression can be extracted from the above equation (\ref{sol r AAdS}).

In this AAdS coordinate representation the metric on the conformal boundary is given by, at the equatorial,
\begin{equation}
 d\bar{s}^2= -dt^2+{1\over l^2}d\Phi^2,
\end{equation}
with conformal factor $Y^2l^2$. On the other hand, the Boyer-Linquist coordinate representation has conformal boundary metric
\begin{equation}
 d\bar{s}'^2= -dt^2+2{a\over \Xi}dt d\phi+{1\over \Xi l^2}d\phi^2,
\end{equation}
with conformal factor $r^2 l^2$. Both conformal boundary metric are related using coordinate transformation
\begin{equation}
 \Phi=\phi-al^2 t
\end{equation}
and so
\begin{equation}
 d\bar{s}'^2={1\over \Xi}d\bar{s}^2.
\end{equation}
It yields that if the CFT theory in $d\bar{s}^2$ has a thermal equilibrium at $T_0$ then it is related, by Tolman's redshifting law~\cite{Tolman:1930zza}, with the CFT theory in $d\bar{s}'^2$ where the temperature has an angular velocity $a$ dependent given by $T(a)=\Xi^{1/2} T_0$. Even further if we consider the full four dimensional metric there will be spatial dependent over $\theta$ which could give different thermodynamical properties of the CFT theory. For more detail on this issue, we refer the reader to consult~\cite{Gibbons:2004ai}.

\subsection{Plots of screening length}
We then compute numerically and plot the screening length for all metrics: (\ref{Boost}), (\ref{Boyer-Linquist}), and (\ref{AAdS}), as functions of angular velocity $\omega$ or $a$. However since we only want to compute the screening length, the maximum of separation length, which is an invariant scalar, it is tempting to expect the plot of, at least, the Kerr-AdS black hole in AAdS coordinates to be similar to the plot of ``Boost-AdS'' black hole case since both asymptotically $AdS_4$ at the boundary\footnote{For the particular case in this paper we may naively use the boost transformation (\ref{boost trans}), though it is not the general symmetry of the Einstein static universe at the boundary, and taking the equatorial slice of the resulting black hole and then compute the screening length.}. In order to make comparison of the plots we need to find relations between all the black hole parameters: AdS curvature $l$, string angular velocity $\omega$, black hole angular velocity $a$, and black hole mass parameters ($M_{Sch}$ and $M_{Kerr}$).
\begin{figure}[ht]
\centerline{\includegraphics[width=8cm]{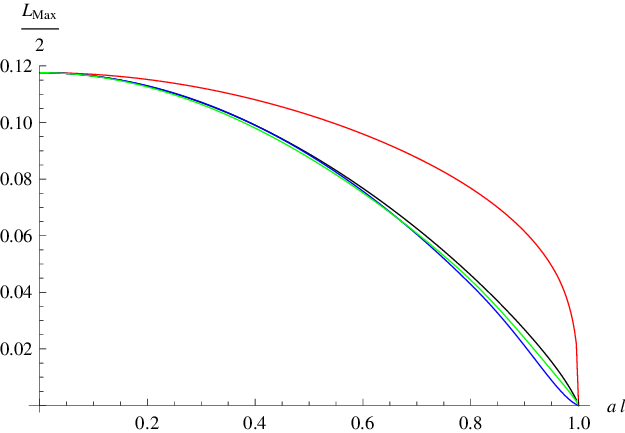}}
\caption{Using software Mathematica we plot numerically the screening length, with $M_{Sch}=M_{Kerr}=30/l$, for different black holes: Schwarzschild-AdS (Black), ``Boost-AdS'' (Red), Boyer-Linquist (Blue), and Asymptotically-AdS (Green). \label{Plots Naive}}
\end{figure}
One immediately notice that all these metrics have the same cosmological constant which is given by $\Lambda=-3l^2$ thus the AdS curvature of all the black holes can be identified to be the same. Angular velocity $\omega$ in (\ref{Boost}) is bounded at boundary by $l^2\geq\omega^2$ while $a$ is also bounded by $a^2 l^2\leq 1$ so we may identify $\omega^2=a^2 l^4$. Remaining parameter that needs to be related is the mass parameters. There is no clear relation between mass parameter of Schwarzschild-AdS black hole, $M_{Sch}$, and mass parameter of Kerr-AdS black hole, $M_{Kerr}$. One could naively take $M_{Sch}=M_{Kerr}$ and try to see if this gives similar physical picture, which is the screening length in our case. As we can see from the ``naive'' plots in Figure \ref{Plots Naive} the screening length of Boyer-Linquist and AAdS black holes are much closer to the screening length of Schwarzschild-AdS black hole rather than to the screening length of ``Boost-AdS'' black hole. Hence it is unlikely that $M_{Sch}= M_{Kerr}$ since we expect that computation of the screening length under all metrics in the rest frame of the string (``Boost-AdS'', Boyer-Linquist, and Asymptotically-AdS) must give a similar behavior.

However we would expect that the mass parameter in the rest frame of quark-antiquark pair should be related to the mass parameter in the rest frame of hot plasma by some power of boost factor,
\begin{align}
 \gamma={1\over\sqrt{1-{\omega^2\over l^2}}}\equiv{1\over\sqrt{1-a^2 l^2}}.
\end{align}
Recall that in the Schwarzschild-AdS black hole the spacetime is static while the Kerr-AdS black hole is the spacetime is stationary. In the rest frame of Schwarzschild-AdS black hole the mass of the black hole is in its rest mass while switching to the reference frame of moving string, the black hole is rotating around the string and so the mass of the black hole now is in its relativistic mass. Therefore we can identify the mass parameter $M_{Sch}$ in the ``Boost-AdS'' black hole is the relativistic mass of the black hole in the view of static string.
\begin{figure}[ht]
\centerline{\includegraphics[width=8cm]{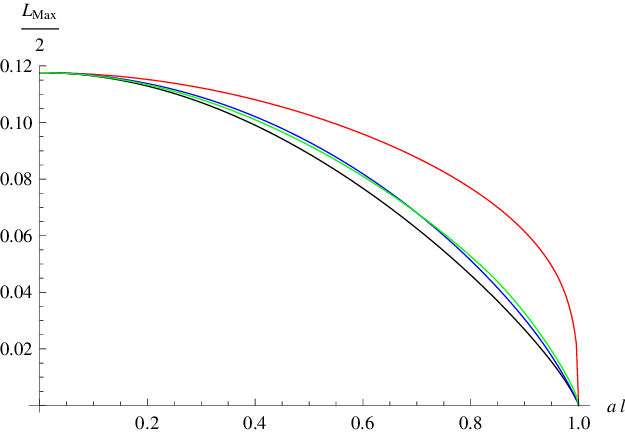}}
\caption{Taking $M_{Sch}=\gamma M_{Kerr}$, we plots the screening length for different black holes: Schwarzschild-AdS (Black), ``Boost-AdS'' (Red), Boyer-Linquist (Blue), and Asymptotically-AdS (Green). \label{Plots Half}}
\end{figure}
On the other hand, based on the result in Figure \ref{Plots Naive}, we could guess that the Kerr-AdS black hole's the mass parameter $M_{Kerr}$ to be static or in its rest mass. So, it is natural to expect considering the usual relativistic effect that $M_{Sch}=\gamma M_{Kerr}$ and the plots are shown in Figure \ref{Plots Half}. One can see that the screening length's plots of Kerr-AdS black hole, in Boyer-Linquist and Asymptotically-AdS coordinate systems, a bit move away from the Schwarzschild-AdS black hole though they are still closer to it than the expected ``Boost-AdS'' black hole.

\begin{figure}[ht]
\centerline{\includegraphics[width=8cm]{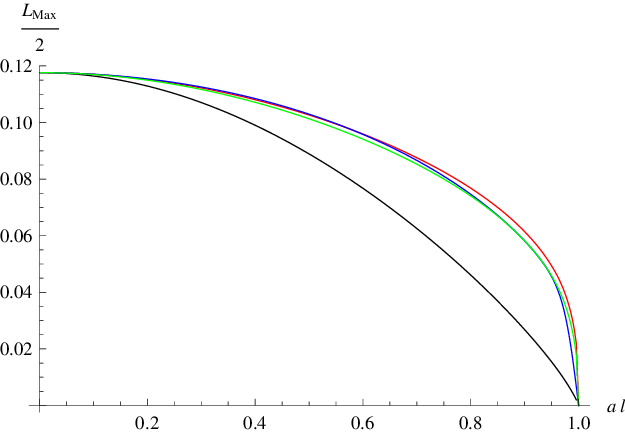}}
\caption{Plots of screening length, with $M_{Sch}=\Xi^{-3/2} M_{Kerr}=30/l$, for different black holes: Schwarzschild-AdS (Black), ``Boost-AdS'' (Red), Boyer-Linquist (Blue), and Asymptotically-AdS (Green). \label{Plots Improved}}
\end{figure}

Using the fact that both ``Boost-AdS'' black hole and Kerr-AdS black hole in Asymptotically-AdS coordinate system are asymptotic to $AdS_4$ we may expect that their string equations of motion should be the same near the boundary, or equivalently the quark-antiquark pair motion. As a result the formula of separation length of both coordinate systems near the boundary should also be the same. Now, suppose that $Y$ is very large providing $l^2Y^2\gg1$ such that $1+l^2Y^2\approx l^2Y^2$. The equation (\ref{sol r AAdS}) now becomes
\begin{align}
\label{large Y}
 \Phi'\approx& {P\Xi^{-1/2} \over l^2Y^4-2M_{Kerr}\Xi^{-3/2}Y}\notag\\
 &\times \sqrt{l^2Y^4-2M_{Kerr}\Xi^{-3/2}Y -a^2 l^4 Y^4 \over l^2Y^4-2M_{Kerr}\Xi^{-3/2}Y-P^2}.
\end{align}
Taking the same large $r$ in (\ref{sol r Boost}), $l^2r^2\gg 1$, we obtain the separation length formula for ``Boost-AdS'' black hole as follows
\begin{align}
\label{large r}
 \phi'\approx {P\gamma \over l^2r^4-2M_{Sch}r}\sqrt{l^2r^4-2M_{Sch}r -a^2 l^4 r^4 \over l^2r^4-2M_{Sch}r-P^2}.
\end{align}
Here we have identified $\omega^2=a^2 l^4$. Near the boundary, we can identify $r=Y$ and compare both formulas (\ref{large Y}) and (\ref{large r}). It turns out that in order for both separation lengths to be equal we must identify
\begin{align}
\label{mass relation}
 M_{Kerr}=M_{Sch}(1-a^2l^2)^{3/2}.
\end{align}
We cheat a bit on the conditions for large $r$ and $Y$ in the computation of separation length formula above. Since we want to keep the mass parameters in the separation formula for large $r$ and $y$, it is a necessary condition to take the mass parameters also very large as such $2M_{Sch}/r\gg1$ and $2M_{Kerr}/Y\gg\Xi^{3/2}$, which means we are considering the plasma with very high temperature. Using relation (\ref{mass relation}) we obtain the numerical plots in Figure \ref{Plots Improved} where now the screening length of Kerr-AdS black holes, Boyer-Linquist and Asymptotically-AdS, closer to the screening length of ``Boost-AdS'' black hole which is in accordance with the physical picture as we expected before.

The unusual mass relation (\ref{mass relation}) is a bit puzzling. Especially when it was considered at the boundary of both Schwarzschild-AdS black hole and Kerr-AdS black hole in AAdS coordinates. A priori there is no relation between the radial coordinate of Schwarzschild-AdS black hole and the radial coordinate of Kerr-AdS black hole in AAdS coordinates on the boundary. One may ask if the mass relation could also give the same physical quantities of the both black holes. One of the important physical quantities is the Hawking temperature which is interpreted as temperature of the plasma at the boundary in the AdS/CFT correspondence prescription. With the killing vector of the metric (\ref{AAdS}) $k=\partial_t+\Omega_K\partial_\Phi$, where $\Omega_K=a(1+l^2Y_K^2)Y_K^{-2}$ is the angular velocity of the black hole relative to the rotating observer at the boundary, the Hawking temperature of Kerr-AdS black hole can be rewritten as
\begin{equation}
 T_{Kerr}={\Xi(1+2l^2Y_K^2)\sqrt{\Xi Y_K^2-a^2}-M_{Kerr} \over 2\pi\Xi Y_K^2}.
\end{equation}
The mass parameter is related to the even horizon by equation $Y_K^2\Xi^2(1+l^2Y_K^2)-2M_{Kerr}\sqrt{\Xi Y_K^2-a^2}=0$. Using the mass relation (\ref{mass relation}), that equation can be simplified to
\begin{equation}
 l^2+{1\over Y_K^2}- {2 M_{Sch} \over Y_K^3}\sqrt{1-{a^2\over \Xi Y_K^2}}=0.
\end{equation}
Now, recall that the horizon in Boyer-Linquist coordinates is related with the horizon in AAdS coordinates by $r_K=\sqrt{\Xi Y_K^2 -a^2}$. In the high Hawking temperature, it gives a condition $\Xi Y_K^2\gg a^2$ and the above equation can be approximated to
\begin{equation}
 l^2+{1\over Y_K^2}- {2 M_{Sch} \over Y_K^3}\approx 0.
\end{equation}
This is the same equation for even horizon in the Schwarzschild-AdS black hole by taking both horizons in Schwarzschild-AdS black hole and Kerr-AdS black hole in AAdS coordinates to be equal, $Y_K=r_H$. Furthermore substituting this to the Hawking temperature of Kerr-AdS black hole, along with the high temperature condition, it turns out to be
\begin{equation}
 T_{Kerr}=\Xi^{1/2}{1+3l^2Y_K^2 \over 4\pi Y_K},
\end{equation}
which is equal to the Hawking temperature of the ``Boost-AdS'' black hole upon identifying $\omega^2= a^2 l^4$. This means that both the CFT theories of the Schwarzschild-AdS black hole and Kerr-AdS black hole in AAdS coordinates indeed have the same temperature which support the mass relation (\ref{mass relation}).

\section{Conclusion and Discussion}
\label{section 4}
We have computed the separation length as a function of momentum transfer $P$ and plotted it for various angular velocity $\omega$. In the plot of Figure \ref{Plots AdS-S}, at each fixed angular velocity there is a lower bound of momentum transfer $P$ which is given by the drag force momentum transfer, $P_c=\omega r_c^2$. At this value the drag force configuration is preferred while below this value there is no physical solution for string configuration of the quark-antiquark pair. For fixed $\omega$, the appearance of this lower bound of non-zero momentum transfer implies that not all separation lengths below the screening length are double value, except for $\omega=0$. Nevertheless the separation length can be divided into two sets identified by $P_c<P<P_{max}$ and $P\geq P_{max}$, in which the later is favored due to its Coulomb potential~\cite{hep-ph/0607062}. 

\begin{figure}[ht]
\centerline{\includegraphics[width=8cm]{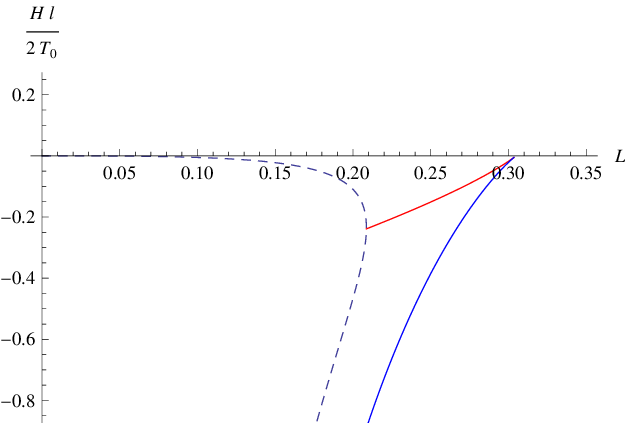}}
\caption{The plot of normalized energy for $\omega=0.5$ where the quark-antiquark pair reaches its screening length at the tip of the plot. \label{Maximum Length}}
\end{figure}

A curious plot shown in the plot of Figure \ref{Plots AdS-S} is the plot of bounded momentum transfer $P=P_c$ colored with black dashed. The profile of this plot is mimicking the profile of separation lengths where it has a maximum ``separation length'' $L=0.104309$ with $P_c=3.98557$, which is the momentum transfer lower bound of plot $\omega/l\approx 0.5$. We have no physical intuition why it is so but if we plot the normalized energy of quark-antiquark pair at angular momentum $\omega=0.5$, as shown in Figure~\ref{Maximum Length}, the separation length in stable region just about reaching its maximum value or the screening length and so we may conclude that this is the minimum angular velocity where the stable quark-antiquark pair could reach its screening length.

Our result and analysis here can be extended to the case of parallel motion of the quark-antiquark pair in the Poincare case where, unlike in the Global case, the distance between quark and antiquark in the pair could take an arbitrary angle about the velocity of the pair or equally the plasma wind, in~\cite{hep-th/0607233} and~\cite{Natsuume:2007vc} it is denoted by $\theta$ which taken value from $\theta=0$ (parallel) to $\theta=\pi/2$ (perpendicular). We can then compare the plots of separation length between the parallel and the perpendicular motion of the quark-antiquark pair. One may wonder if the parallel plot is a result from the perpendicular plot after removing the lines below the plot of bounded momentum transfer. A quick observation of Figure \ref{Plots AdS-S} shows that these plots may not be part of the perpendicular plots. Indeed, the full parallel plot was computed in~\cite{Argyres:2006vs} for five dimensional AdS-black hole in Poincare case and one can see it is different from the perpendicular plot. Here, we are reluctant to compute the full plot since the quark-antiquark pair does not exist physically below the plot of momentum transfer as discussed in section {\ref{section 2}. It was shown in~\cite{Natsuume:2007vc} that, for fixed velocity, the parallel plot has higher screening length than the perpendicular plot. To understand this intuitively assume that there is a continuous deformation of separation length in terms of the angle $\theta$ we then may consider the plot of bounded momentum transfer in the perpendicular plot as a horizontal line $L=0$. We would expect to see a slowly formation of the plot of bounded momentum transfer from a horizontal line in the perpendicular plot to the dashed line in the parallel plot. Decreasing the $\theta$, we would expect that the plots of separation length will be shifted up from its corresponding perpendicular curves. As a result the screening length for each non-zero fixed velocity in any value of $\theta$ is always bigger than the corresponding screening length in perpendicular plot, $\theta=\pi/2$, except for zero velocity plot which is unchanged.

The existence of bounded momentum transfer is puzzling. We could ask if there a continuous transformation from the quark-antiquark configuration to the drag force configuration. As one can see from Figure \ref{Plots AdS-S} close to the $P_c$ the separation length value is below the screening length however at $P=P_c$ there is a sudden jump in the separation distance between the quark and the antiquark where each of them considered as in drag force configuration with the distance between them is much larger than the screening length, as a process shown in Figure \ref{string conf} and followed by discussion before the subsection \ref{drag force}. However we could simply avoid this puzzle since this possible continuous transformation happens in the metastable region of the plots in Figure \ref{Plots AdS-S} which is not physical. 

Unlike the perpendicular case, there is a drag force experience by the quark-antiquark pair as discussed in section \ref{drag force}. As shown in Figure~\ref{Plots AdS-S}, we would expect that the drag force in the metastable region would be similar to the one derived from the drag force configuration for at least in the non-relativistic limit~\cite{NataAtmaja:2010hd}. This is likely to be true for the metastable region, but not for the stable region, because the tip of the string configuration is closer to the horizon compared to the stable region and so the string still feel the presence of plasma. Another reason is because string configuration of the metastable region is also close to the drag force configuration for $P\gtrsim P_c$ and so it is expected $P\approx \gamma~p_\phi$, where $\gamma$ is a constant. In the stable region the plots are very dense in the non-relativistic limit thus the momentum transfer, although very large, may not depend on the angular velocity and thus it can not be interpreted as drag force.

Besides separation length of the quark-antiquark pair, we also computed the bounded energy density of the quark-antiquark pair. The bounded energy density is divergent at the boundary. This divergence can be removed by subtracting the individual quark and antiquark energy densities. However this cost a new divergence at the horizon and to resolve this problem we need to go to the reference frame of the moving quark-antiquark pair where the background metric is ``Boost-AdS'' black hole. We also found that all the normalized bounded energy of quark-antiquark pair has minimum length in the metastable region bounded by a dashed plot shown in Figure \ref{fig:dummy}.

The demand of going to the reference frame of quark-antiquark pair we obtained a ``Boost-AdS'' black hole. However there is also known stationary solution of the AdS black hole called Kerr-AdS black hole. A priori there is no relation between parameters of both Kerr-AdS black hole and ``Boost-AdS'' black hole. As discussed in section \ref{section 3} we argued that some of these parameters are related. In particular we computed the screening length of the static quark-antiquark pair under these black holes and compared with the previous calculation of Schwarzschild-AdS black hole. We concluded that $M_{Kerr}=M_{Sch}(1-a^2l^2)^{3/2}$ is more suitable in our physical picture especially for the high temperature plasma. In fact, we could just use the condition at the horizon and, with the range of radial coordinate in AAdS is $0<Y_K\leq Y<\infty$, we could take for any radius that $\Xi Y^2\gg a^2$ is valid. Using this the metric of ``Boost-AdS'' black hole (\ref{Boost}) is equal to the metric of Kerr-AdS black hole (\ref{AAdS}) providing that we identify the both radius coordinates to be the same and use all the relation of black hole parameters given previously. This means that the equilibrium rotating plasma of the CFT theory is effectively the same in the equatorial for both metrics: ``Boost-AdS'' black hole (\ref{Boost}) or the Kerr-AdS black hole (\ref{AAdS}). However this is only valid at the equatorial and we could ask whether this is also valid in general for arbitrary polar angle $\theta$.

Recall that transforming the Schwarzschild-AdS black hole using transformation (\ref{boost trans}) gives a metric that is still the solution of four dimensional Einstein equation with negative cosmological constant. As the Hawking temperature is still $T_{Boost}$, it would be interesting to study the thermodynamical properties of this metric compared to the Kerr-AdS black hole. If all thermodynamical properties are equal hence it is better to work in this metric rather than working in complicated AAdS coordinate representation of the Kerr-AdS black hole. If not, we suspect there might be a coordinates transformation, that depend on polar angle $\theta$, which is also a symmetry of three dimensional Einstein universe. Applying this to the Schwarzschild-AdS black hole gives a metric that could still be a solution to the Einstein equation with negative cosmological constant. The thermodynamical properties of this metric might be comparable to the Kerr-AdS black hole although the metric might not be simple.

In the literature, we have not found so far the relation between mass parameter of Schwarzschild-AdS black hole $M_{Sch}$ and mass parameter of Kerr-AdS black hole $M_{Kerr}$. A general procedure for constructing the metric of Kerr-AdS black holes in~\cite{Gibbons:2004uw} was built out of the AdS metric plus additional terms, scaled with mass parameter of Kerr-AdS, in the Kerr-Schild form. In that sense the mass parameter is added by hand thus it has no clear direct connection with the mass parameter of Schwarzschild-AdS black hole. It would be interesting to construct the Kerr-AdS black hole from the Schwarzschild-AdS black hole, e.g. by following the Newman-Janis procedure~\cite{Newman1965}, and to check if our the mass relation obtained in this article is valid.

%

\begin{acknowledgements}
A.N.A. is grateful to KEK group theory for hospitality and in particular to Makoto Natsuume for useful discussion during the initial work of this article. A.N.A, H.A.K. and N.Y. acknowledge University of Malaya for the support through the University of Malaya Research Grant (UMRG) Programme RP006C-13AFR and RP012D-13AFR.
\end{acknowledgements}


\begin{thebibliography}{}

\bibitem{BNL-38344} 
  T.~Matsui and H.~Satz,
  ``J/psi Suppression by Quark-Gluon Plasma Formation,''
  Phys.\ Lett.\ B\ {\bf 178}, 416  (1986).

\bibitem{Callan:1997kz} 
  C.~G.~Callan and J.~M.~Maldacena,
  ``Brane death and dynamics from the Born-Infeld action,''
  Nucl.\ Phys.\ B {\bf 513}, 198 (1998)
  [hep-th/9708147].

\bibitem{hep-th/9803002} 
  J.~M.~Maldacena,
  ``Wilson loops in large N field theories,''
  Phys.\ Rev.\ Lett.\ \ {\bf 80}, 4859  (1998)
  [hep-th/9803002].

\bibitem{hep-th/9803001}
  S.~J.~Rey and J.~T.~Yee,
  ``Macroscopic strings as heavy quarks in large N gauge theory and  anti-de
  Sitter supergravity,''
  Eur.\ Phys.\ J.\  C {\bf 22}, 379 (2001)
  [arXiv:hep-th/9803001].

\bibitem{hep-th/9803135} 
  S.~-J.~Rey, S.~Theisen and J.~-T.~Yee,
  ``Wilson-Polyakov loop at finite temperature in large N gauge theory and anti-de Sitter supergravity,''
  Nucl.\ Phys.\ B\ {\bf 527}, 171  (1998)
  [hep-th/9803135].

\bibitem{Brandhuber:1998bs} 
  A.~Brandhuber, N.~Itzhaki, J.~Sonnenschein and S.~Yankielowicz,
  ``Wilson loops in the large N limit at finite temperature,''
  Phys.\ Lett.\ B {\bf 434}, 36 (1998)
  [hep-th/9803137].

\bibitem{hep-ph/0607062} 
  H.~Liu, K.~Rajagopal and U.~A.~Wiedemann,
  ``An AdS/CFT Calculation of Screening in a Hot Wind,''
  Phys.\ Rev.\ Lett.\ \ {\bf 98}, 182301  (2007)
  [hep-ph/0607062].

\bibitem{hep-th/0607233} 
  E.~Caceres, M.~Natsuume and T.~Okamura,
  ``Screening length in plasma winds,''
  JHEP\ {\bf 0610}, 011  (2006)
  [hep-th/0607233].

\bibitem{hep-th/0607089} 
  M.~Chernicoff, J.~A.~Garcia and A.~Guijosa,
  ``The Energy of a Moving Quark-Antiquark Pair in an N=4 SYM Plasma,''
  JHEP\ {\bf 0609}, 068  (2006)
  [hep-th/0607089].
%

\bibitem{Becattini:2007sr}
  F.~Becattini, F.~Piccinini and J.~Rizzo,
  ``Angular momentum conservation in heavy ion collisions at very high
  energy,''
  Phys.\ Rev.\  C {\bf 77}, 024906 (2008)
  [arXiv:0711.1253 [nucl-th]].

\bibitem{Abreu:2007kv}
  N.~Armesto {\it et al.},
  ``Heavy Ion Collisions at the LHC - Last Call for Predictions,''
  J.\ Phys.\ G {\bf 35}, 054001 (2008)
  [arXiv:0711.0974 [hep-ph]].

\bibitem{Csernai:2011qq} 
  L.~P.~Csernai, D.~D.~Strottman and C.~Anderlik,
  ``Kelvin-Helmholz instability in high energy heavy ion collisions,''
  Phys.\ Rev.\ C {\bf 85}, 054901 (2012)
  [arXiv:1112.4287 [nucl-th]].

\bibitem{Wang:2013xtp} 
  D.~J.~Wang, Z.~Néda and L.~P.~Csernai,
  ``Viscous potential flow analysis of peripheral heavy ion collisions,''
  Phys.\ Rev.\ C {\bf 87}, no. 2, 024908 (2013)
  [arXiv:1302.1691 [nucl-th]].

\bibitem{McInnes:2014haa} 
  B.~McInnes,
  ``Angular Momentum in QGP Holography,''
  arXiv:1403.3258 [hep-th].
  
\bibitem{Gibbons:2004ai}
  G.~W.~Gibbons, M.~J.~Perry and C.~N.~Pope,
  ``The first law of thermodynamics for Kerr - anti-de Sitter black holes,''
  Class.\ Quant.\ Grav.\  {\bf 22}, 1503 (2005)
  [arXiv:hep-th/0408217].
  
\bibitem{Hemming:2007yq}
  S.~Hemming and L.~Thorlacius,
  ``Thermodynamics of Large AdS Black Holes,''
  JHEP {\bf 0711}, 086 (2007)
  [arXiv:0709.3738 [hep-th]].

\bibitem{Gubser:2006bz}
  S.~S.~Gubser,
  ``Drag force in AdS/CFT,''
  Phys.\ Rev.\  D {\bf 74}, 126005 (2006)
  [arXiv:hep-th/0605182].

\bibitem{Herzog:2006gh}
  C.~P.~Herzog, A.~Karch, P.~Kovtun, C.~Kozcaz and L.~G.~Yaffe,
  ``Energy loss of a heavy quark moving through N = 4 supersymmetric
  Yang-Mills plasma,''
  JHEP {\bf 0607}, 013 (2006)
  [arXiv:hep-th/0605158].

\bibitem{NataAtmaja:2010hd}
  A.~Nata Atmaja and K.~Schalm,
  ``Anisotropic Drag Force from 4D Kerr-AdS Black Holes,''
  JHEP {\bf 1104}, 070 (2011)
  [arXiv:1012.3800 [hep-th]].

\bibitem{Chunlen:2010ar}
  S.~Chunlen, K.~Peeters and M.~Zamaklar,
  ``Finite-size effects for jet quenching,''
  arXiv:1012.4677 [hep-th].

\bibitem{Natsuume:2007vc} 
  M.~Natsuume and T.~Okamura,
  ``Screening length and the direction of plasma winds,''
  JHEP {\bf 0709}, 039 (2007)
  [arXiv:0706.0086 [hep-th]].

\bibitem{Argyres:2006vs} 
  P.~C.~Argyres, M.~Edalati and J.~F.~Vazquez-Poritz,
  ``No-drag string configurations for steadily moving quark-antiquark pairs in a thermal bath,''
  JHEP {\bf 0701}, 105 (2007)
  [hep-th/0608118].
  
\bibitem{Bhattacharyya:2008jc} 
  S.~Bhattacharyya, V.~EHubeny, S.~Minwalla and M.~Rangamani,
  ``Nonlinear Fluid Dynamics from Gravity,''
  JHEP {\bf 0802}, 045 (2008)
  [arXiv:0712.2456 [hep-th]].
  
\bibitem{Carter:1968ks} 
  B.~Carter,
  ``Hamilton-Jacobi and Schrodinger separable solutions of Einstein's equations,''
  Commun.\ Math.\ Phys.\  {\bf 10}, 280 (1968).

\bibitem{Hawking:1998kw}
  S.~W.~Hawking, C.~J.~Hunter and M.~Taylor,
  ``Rotation and the AdS/CFT correspondence,''
  Phys.\ Rev.\  D {\bf 59}, 064005 (1999)
  [arXiv:hep-th/9811056].

\bibitem{Kim:1998iw} 
  H.~Kim,
  ``Spinning BTZ black hole versus Kerr black hole: A Closer look,''
  Phys.\ Rev.\ D {\bf 59}, 064002 (1999)
  [gr-qc/9809047].

\bibitem{Newman1965}
E. T. Newman and A. I. Janis,``Note on the Kerr Spinning Particle Metric,''
Journal of Mathematical Physics {\bf 6}, 915-917 (1965).
  
\bibitem{Gibbons:2004uw}
  G.~W.~Gibbons, H.~Lu, D.~N.~Page and C.~N.~Pope,
  ``The general Kerr-de Sitter metrics in all dimensions,''
  J.\ Geom.\ Phys.\  {\bf 53}, 49 (2005)
  [arXiv:hep-th/0404008].
  
\bibitem{Witten98-1}
  E.~Witten,
  ``Anti-de Sitter space and holography,''
  Adv.\ Theor.\ Math.\ Phys.\  {\bf 2}, 253 (1998)
  [arXiv:hep-th/9802150].

\bibitem{Gubser:1998bc}
  S.~S.~Gubser, I.~R.~Klebanov and A.~M.~Polyakov,
  ``Gauge theory correlators from non-critical string theory,''
  Phys.\ Lett.\  B {\bf 428}, 105 (1998)
  [arXiv:hep-th/9802109].

\bibitem{Tolman:1930zza} 
  R.~C.~Tolman,
  ``On the Weight of Heat and Thermal Equilibrium in General Relativity,''
  Phys.\ Rev.\  {\bf 35}, 904 (1930).

\end{thebibliography}


\end{document}